\documentclass{article}
\usepackage{natbib}
\usepackage{amsmath} 
\usepackage{ascmac} 

\usepackage{xcolor}
\usepackage[a4paper, top=24truemm,bottom=29truemm,left=25truemm,right=25truemm]{geometry}
\usepackage{graphicx}

\title{Mitigation of spatial economic impact propagation of highway disruptions by redundant networks}
\author{Tomoki Ishikura\thanks{Tokyo Metropolitan University, Minamiosawa 1-1, Hachioji, Tokyo, 192-0397 Japan
e-mail: iskr@tmu.ac.jp}
}

\usepackage[bookmarks=false]{hyperref}
    \hypersetup{colorlinks,
      linkcolor=blue,
      citecolor=blue,
      urlcolor=blue}

\begin{document}






\maketitle

\begin{abstract}
The damage to transportation infrastructure caused by disasters can indirectly lead to economic damage through economic interdependence, even in areas that are not directly affected.
However, even when transportation routes are interrupted by a disaster, the damage can be mitigated if alternative routes are secured.
Rural areas with low-density transportation networks are more vulnerable to traffic disruptions in a disaster.
This study develops a method for evaluating the effectiveness of redundant transportation networks in mitigating economic vulnerability in the event of a disaster.
Our methodology combines inter-regional road network connectivity with a spatial computable general equilibrium (SCGE) model.
We apply the method to road disruption scenarios in the Chugoku region of Japan, which has a system of parallel highways.
The affected areas are in close geographical proximity to many rural areas and have strong economic interdependencies with them.
Several counterfactual simulations depicted the situation without the alternative road and the disaster.
We evaluate the transportation impacts, measured by changes in travel time, and the economic impacts, measured by negative benefits, respectively.
The results suggest that the economic vulnerability reduction effect is more far-reaching than the transportation impacts.

\end{abstract}


\section{Introduction}
\label{introduction}

Damage to transportation infrastructure from disasters not only causes human and property losses but also indirectly impacts economic damage in areas not directly affected by the disaster through the interdependence of socioeconomic activities.
In the short term, the disruption of transportation networks may obstruct rescue activities and cause socioeconomic isolation in the affected areas.
In the medium to long run, increasing inter-regional transportation costs (generalized costs) may lead to increased production costs and decreased demand for goods and services.
The region would be economically vulnerable if the performance of inter-regional trade depended on such critical roads.

However, when a disaster disrupts a certain transportation route, the damage is mitigated if alternative routes are available.
This characteristic is called the redundancy (or multiplicity) of transportation networks.
\cite{hakusho2012E} emphasizes the importance of redundancy by referring to the example of the Great East Japan Earthquake, which damaged many transportation networks, and the fact that the unaffected routes performed a significant role as detour routes and emergency transportation routes.

Nowadays, climate change has caused more frequent and severe weather hazards such as typhoons and torrential rains.
In Japan, the Nankai Trough Megathrust Earthquake is expected to occur shortly, bringing a massive disaster to the southern coast of Japan.
It means the Japanese economy undoubtedly faces the risk of disasters disrupting transportation systems.
Evaluating the effectiveness of network redundancy to mitigate the impact of such disasters, namely economic vulnerability, is critical for disaster preparedness policy.

This study develops a framework for evaluating the effectiveness of redundant transportation networks in mitigating economic vulnerability in the event of a disaster.
Specifically, we propose a concept to measure the effect of specific road corridors on network redundancy.
A key feature of our approach is separating the impact of network disruption on transportation and the economic impact.

The remainder of this paper is organized as follows.
Section 2 is the literature review, and it is followed by the theoretical framework of the methodology in Section 3.
Section 4 presents case studies and discusses the characteristics of the results. 
A final section concludes the paper.

\section{Literature review}
\label{review}

Quantitative evaluation of the effects of network redundancy has focused much attention in recent years.
\cite{xu2021} presented a methodology for increasing the resilience of transportation networks by optimizing route diversity and transportation network redundancy.
For the value of road network redundancy, \cite{jenelius2015} proposed a method to evaluate the value of road network redundancy using the reduction of welfare losses of transportation users as a measure and applied it to public transportation in Stockholm.
\cite{zhu2023} develops a framework to investigate the economic feasibility of adding redundancy to road networks by introducing benefit-cost ratio (BCR) for the stochastic disruption scenarios and applies it to the Winnipeg network.
These studies focus on the benefits for transportation users, mainly assuming an urban transportation network.
The geographic scope of the users affected by a traffic disruption is assumed to be within the urban area.

When intercity highways are disrupted, the affected transportation users are not limited to the directly affected area, but spread out over a wider area.
Furthermore, the economic impact spillover extends not only to transportation users but also to industrial activities.
The impact on freight traffic will influence the cost of goods production in other regions through the supply chain.
This means that the impacts of road disruptions also affect producers and households in areas that do not use that road section.
Such indirect effects have not been considered in studies that focus only on the benefits of transportation users.
\cite{okuyama2019} investigated a comprehensive review of the various methodologies for analyzing the economic damage spillovers from disasters.

The integrated frameworks of transportation network and economic analysis have been applied to the spatial economic impact assessment of disasters.
\cite{kim2002} and \cite{ishikura2024} analyze the regional economic and interregional trade impacts of floods and volcanic eruptions, respectively, using an integrated approach of transportation network model and input-output model.
\cite{cho2015} developed a model that integrates a transportation network equilibrium model and a multi-regional input-output model to internalize changes in final demand resulting from changes in transportation costs.
\cite{cho2015} applied the model to several hypothetical scenarios of highway disruptions in the United States, evaluating changes in transportation costs between origin-destination pairs and their economic impacts.
Nevertheless, such models based on Leontief-type production technology overlook the substitution of inputs in response to changes in the prices of goods.
In other words, these methodologies dismiss the impact of transportation disruptions on the production cost structure of industrial sectors.

\cite{rose2005} applied a computable general equilibrium (CGE) model to the case of water supply disruptions to evaluate the resilience of regional economies to disasters.
\cite{kajitani2018} and \cite{yang2023} explored the inter-regional and inter-industry spillovers of economic damage from earthquakes using the Spatial CGE (SCGE) models, which explicitly consider the generalized transportation costs.


However, few studies have quantitatively evaluated the effects of network redundancy by considering inter-regional spillovers (\cite{faturechi2015}).
This paper presents a methodology to measure the impact of the presence or absence of alternative routes on a multi-regional economic system in the event of transportation disruptions due to a disaster, and to evaluate the regional effects of network redundancy.
Furthermore, we apply this methodology to a highway disruption during the July 2018 heavy rain event in western Japan, and explore the contribution to redundancy of the existence of the Chugoku Expressway, which was focused on as an alternative route at the time.

\section{Methodology}
\label{methodology}

When a disaster disrupts the road sections, the generalized cost of interregional trade using those sections as a transportation route increases.
If other routes can serve as alternatives, the negative impact on the regional economy will be mitigated.
Such a network with diversified routes is called a redundant network and can contribute to mitigating the economic vulnerability of the regions, which is considered a component of disaster resilience.

There is no commonly accepted quantitative definition of transportation network redundancy.
For example, \cite{xu2021} quantitatively evaluated the redundancy using the diversity of travel routes as an index.
When dealing with inter-city trunk transportation networks, the diversity of available routes may not provide an appropriate measure.
Thus, the disruption of a road section will affect the incremental transportation costs of using alternative routes more than the diversity of alternative routes.

This study considers the extent to which the presence of a road system mitigates the damage caused by road disruptions, contributing to the redundancy of the network.
To investigate the contribution to the redundancy, we define four types of states.

We assume a normal state (N) and a state in which some road sections are disrupted by a disaster (D).
For the state of the transportation network, we assume an actual road network (A) and a counterfactual network in which the road section to be assessed does not exist (C).
The combination of these two states allows us to represent the state of the road network as four different states (Table \ref{tab:4 state of network}).

The difference between these four states defines the following variations.
\begin{enumerate}
	\item Disaster damage in the actual network: the difference between A-N and A-D
	\item Disaster damage in counterfactual network: the difference between C-N and C-D
	\item Effect of installation of the section: Difference between A-N and C-N
	\item Effect of network redundancy on disaster mitigation by the section: Difference between 1 and 2
\end{enumerate}
Disaster damage is the direct impact of a road disruption caused by a disaster.
In both the actual and counterfactual networks, the disruption of a damaged section will change the travel time between OD pairs that include it in the shortest route.

The installation effect is the difference in transportation and economic performance with and without the section to be evaluated.
This is equivalent to the concept of standard transportation investment effects.

We define the effect of network redundancy as the difference between the disaster damage in the actual network and the disaster damage in the counterfactual network.
If the section to be evaluated serves as alternative paths, the impact of the disaster is mitigated in some regions.

\begin{table}
	\centering
	\caption{Assumption of network state}
	\label{tab:4 state of network}
	\vspace{5mm}
	\begin{tabular}{c|c c}
		& normal state (N) & damaged state (D)\\ \hline
		\begin{tabular}{c}
			actual network\\
			(A)
		\end{tabular} & A-N & A-D \\
		\begin{tabular}{c}
			counterfactual network\\
			(C)
		\end{tabular} & C-N & C-D \\ \hline
	\end{tabular}
\end{table}

The transportation effect is measured in terms of vehicle travel time.
Aggregating the minimum travel time to all other regions $t^S_{ij}$ per origin or destination $i$ is regarded as the road accessibility index of the region $i$, $T^S_i$
\begin{equation}
	\label{definition: accessibility}
	T^S_i = \sum_{j \ne i} t^S_{ij},
\end{equation}
where $S$ denotes state of the network and $i(j)$ denotes the region.

Using the accessibility index, the disaster damage $ZT^S_i$ at network $N$ in region $i$ is defined as follows
\begin{align}
	\label{disaster damage, transport}
	ZT_i^A &= T^{A-D}_i - T^{A-N}_i,\\
	ZT_i^C &= T^{C-D}_i - T^{C-N}_i.
\end{align}
Meanwhile, the effect of network redundancy in region $i$ is defined as 
\begin{align}
	\label{reduncandy effect, transport}
	RT_i & = ZT_i^C - ZT_i^A \notag \\
	& = \left( T^{C-D}_i - T^{C-N}_i\right)
		=\left( T^{A-D}_i - T^{A-N}_i \right).
\end{align}
The difference in the accessibility index between the network states under normal conditions $T^{A-N}_i - T^{C-N}_i$ indicates the effect of the installation of the relevant section.

Changes in traffic conditions imply changes in inter-regional transportation costs.
Therefore, focusing on changes in economic states, changes in transportation conditions shift equilibrium prices and equilibrium demand.
Concerning the economic evaluation, our method focuses on the welfare level, which is a consequence of the general equilibrium.

\begin{figure}[t]
	\begin{minipage}{0.45\columnwidth}
		\centering
		\includegraphics[width=0.8\linewidth]{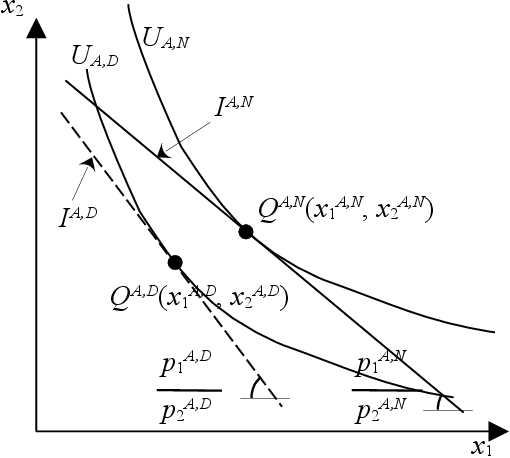}
		\caption{Shift of equilibrium state by a disaster}
		\label{fig:equilibrium change}
	\end{minipage}
	\hspace{0.04\columnwidth} 
	\begin{minipage}{0.45\columnwidth}
		\centering
		\includegraphics[width=0.8\linewidth]{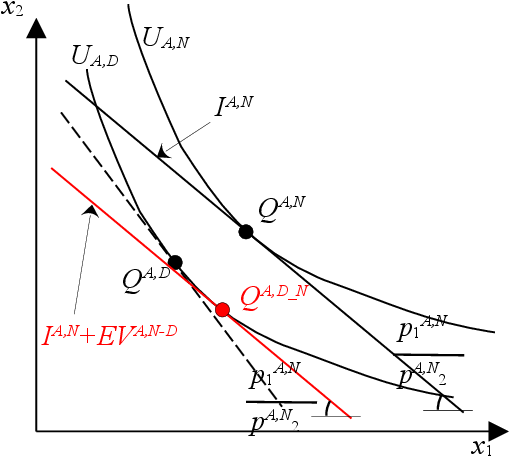}
		\caption{Equivalent variation, (A-N)-(A-D)}
		\label{fig:concept EV}
	\end{minipage}
\end{figure}

Suppose the equilibrium shifts in the two-goods market as shown in the Fig. \ref{fig:equilibrium change}.
The economy under normal conditions in the actual network (A) is in equilibrium at the point of utility-maximizing goods consumption $Q^N(x_1^{A,N},x_2^{A,N})$ under the price vector $(p_1^{A,N},p_2^{A,N})$.
The equilibrium expenditure in the state is $I^{A,N}N$ and the utility level is $U_{A,N}$.
When a disaster disrupts certain transportation sections, the generalized cost of transportation services using the sections will increase.
As a result, the equilibrium state will shift to $Q^{A,D} (x_1^{A,D},x_2^{A,D})$ at the new price system $(p_1^{A,D},p_2^{A,D})$.
We now sppose that $p_1$ is significantly affected directly by the disaster $(p_1^{A,D} \gg p_1^{A,N})$.
Assuming no significant change in the income level, the increase in the price of transportation services due to the disaster generally shifts the budget constraint line downward, and the feasible utility level lowers to $U_{A,D}$.

Measuring this difference in utility levels in monetary terms evaluates the benefit as a measure of welfare change.
However, the monetary value can vary depending on the price system of the states based on measurement.
Of the four equilibria in Table \ref{tab:4 state of network}, A-N is the benchmark equilibrium state that can be observed, while all other states are estimated economic equilibria.
Therefore, the equilibrium state A-N may be the depiction with the least uncertainty, such as estimation error.
We use the equivalent variation (EV) index based on the price system in the A-N equilibrium, which is the state before any changes occur in the economy.

EV is defined as the amount of additional income needed to get the same level of utility they could have reached if the economic conditions had changed, which is measured in the price system of the state before the change in conditions (Fig. \ref{fig:equilibrium change}).
Using the expenditure function $e(U,p_1,p_2)$, the economic value of disaster damage $EV^{A,N-D}$ in the actual network is defined as the difference between the minimum expenditure of $Q^{A,N}$ and $Q^{A,D}$ as follows.
In general, $EV$ will be negative if the utility level is worsened by a disaster.

Likewise, the difference in welfare between the real network situation (A) and the counterfactual network situation (C) in the normal state without disaster (N) can be defined by $EV^{N,A-C}$, the income required to reach utility in (C) in the (A) price range system.
This is equivalent to the compensating variation evaluated in the (A) price system for the change from state ($C$) to state ($A$) by transportation improvements.

In our approach, comparisons including ($C$-$D$) states require a particular interpretation.
These comparisons can be directly evaluated as EV or CV using an expenditure function assessed at either the C-N or C-D price system.
However, none of these are directly comparable to the (negative) benefits of disaster damage in actual networks, as measured by EV indicators based on the A-N price system.
To measure the effects of network redundancy, we need to unify the A-N pricing system as the basis for monetary values.

\begin{figure}
	\begin{minipage}{0.45\columnwidth}
		\centering
		\includegraphics[width=0.8\linewidth]{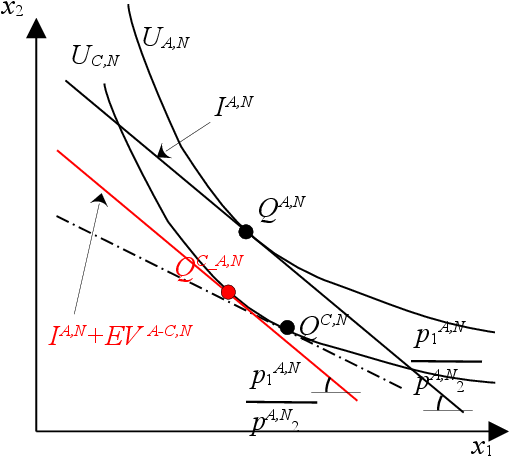}
		\caption{Equivalent variation, (A-N)-(C-N)}
		\label{fig:concept counterfactual EV}
	\end{minipage}
	\hspace{0.04\columnwidth} 
	\begin{minipage}{0.45\columnwidth}
		\centering
		\includegraphics[width=0.8\linewidth]{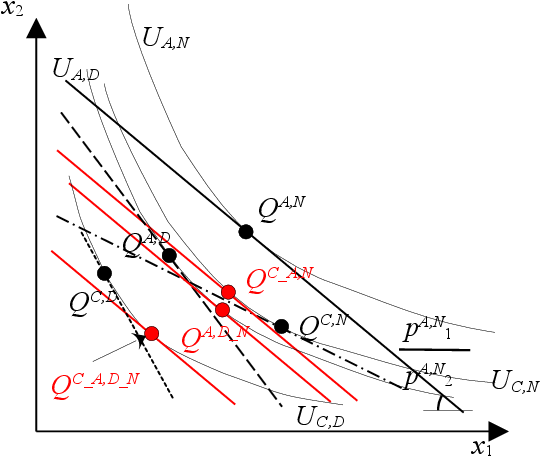}
		\caption{Utility level based on (A-N) price system}
		\label{fig:concept all EV}	
	\end{minipage}
\end{figure}

For this reason, even for effects that include C-D states in the comparison, we use the expenditure function based on the A-N price system to evaluate the benefit.
Specifically, Fig. \ref{fig:concept all EV} explains the evaluation concept.
The general equilibrium model derives equilibrium states $Q^{A,N},Q^{A,D},Q^{C,N},Q^{C,D}$ and corresponding utility levels $U_{A,N},U_{A,D},U_{C,N},U_{C,D}$ for A-N, A-D, C-N and C-D states respectively.
Assuming that these utility levels are achieved under the A-N price system, $Q^{A,N},Q^{A,D_N},Q^{C_A,N},Q^{C_A,D_N}$, the difference between these budget constraints represents the benefit in the unified value term.
That is, the difference in the evaluated value of the expenditure function corresponding to each equilibrium represents the EV for the utility change from A-N, based on the A-N price system.
Using this concept, we define the disaster damage in the counterfactual network as “quasi-EV” ($qEV^{C,N-D}$),
\begin{equation}
	\label{eq: quasi EV counterfactual net}
	qEV^{C,N-D} = e\left(U_{C,D},p_1^{A,N},p_2^{A,N}\right) - e\left(U_{C,N},p_1^{A,N},p_2^{A,N}\right).
\end{equation}

Therefore, $R$, the effect of the redundant network due to the existence of alternative routes, is the difference between $EV^{A,N-D}$, the disaster economic damage in the network (A), and $qEV^{C,N-D}$, the disaster economic damage in the network (C).
\begin{align}
	\label{eq: effects of redundancy}
	R	& = EV^{A,N-D} - qEV^{C,N-D}\notag \\
		& = e\left(U_{A,D},p_1^{A,N},p_2^{A,N}\right) - e\left(U_{A,N},p_1^{A,N},p_2^{A,N}\right)
		- e\left(U_{C,D},p_1^{A,N},p_2^{A,N}\right) - e\left(U_{C,N},p_1^{A,N},p_2^{A,N}\right)
\end{align}

\section{Case study: a highway section disruption of the Sanyo Expressway in Hiroshima Japan}
\label{Case study}
\subsection{Assumption of the states and data}
\label{assumptions}

This paper takes the partial disruption of the Sanyo Expressway for the case studies.
The road section was closed due to a landslide caused by the Heavy Rain Event in July 2018.
The landslides occurred simultaneously over a wide area, and many major roads were closed for several days \footnote{Chugoku Regional Development Bureau, https://www.cgr.mlit.go.jp/photo/h3007gouu\_kiroku/index.htm, accessed 28 October 2024} in July 2018.
The disaster caused extensive damage not only to the Sanyo Expressway but also to the entire Chugoku region (west side of the Japanese main island).

Although some sections of the Sanyo Expressway, the major trunk road of the east-west corridor in the region, were shut down, two alternative routes (the Chugoku Expressway and the San'in Expressway) greatly contributed to the preservation of wide-area transportation functions.
The volume of traffic on the Chugoku Expressway was approximately five times that of normal traffic due to the inflow of traffic on the damaged road.
This phenomenon suggests that the Chugoku Expressway served as an alternative route to the disrupted road.

\begin{figure}
	\centering
	\includegraphics[width=0.7\linewidth]{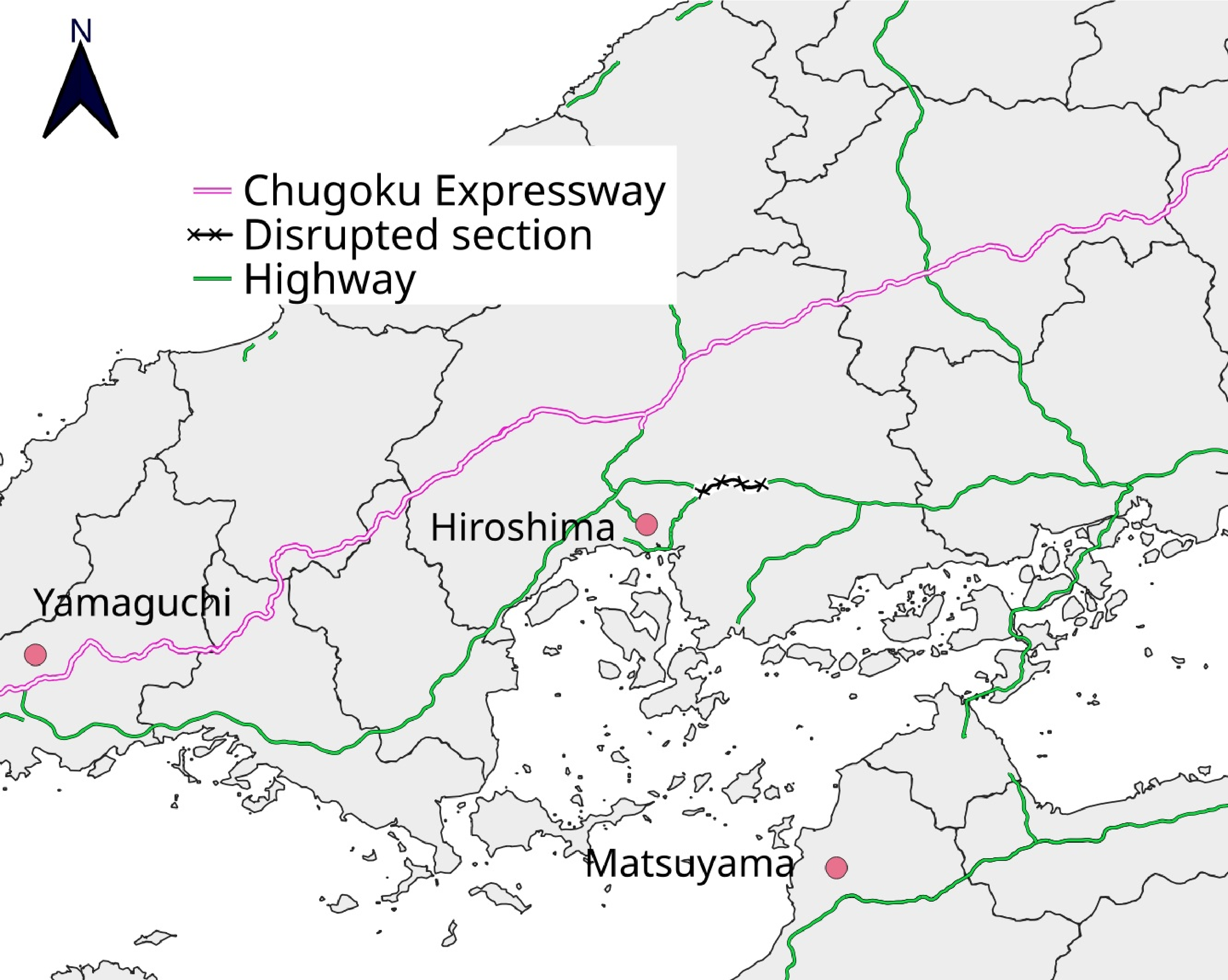}
	\caption{Disrupted link and Chugoku Expressway}
	\label{fig:setting map}
\end{figure}

Although the disruption lasted less than ten days at that time, which is a relatively short period in the context of shifting economic equilibrium, the road section in question runs through a high-risk area for landslides so that future hazards could result in more prolonged disruptions.
We evaluate the contribution of the Chugoku Expressway, which lies parallel to the disrupted section, as an alternative route to enhance network redundancy.
The geographical image of the area is shown in Fig. \ref{fig:setting map}.

We apply the proposed method to quantitatively estimate the effect of the redundant network.
This case study regards a hypothetical road network, in which only the Chugoku Expressway is removed from the actual expressway system, as the counterfactual.
We estimate the traffic and economic status of the real network under normal and disaster conditions, and the counterfactual network under normal and disaster conditions, respectively, and discuss the effects of the redundant network.
The road network state corresponding to (D) and (C) in Table \ref{tab:4 state of network} is assumed as shown in Table \ref{tab: case study state}.

\begin{table}
	\centering
	\caption{Assumption of the case study}
	\label{tab: case study state}
	\vspace{5mm}
	\begin{tabular}{c c}
		assumed state & description \\
		\hline damaged state (D) & 
			\begin{tabular}{c}
				A section disrupted \\
				(Hiroshima-Higashi IC - Shiwa IC)
			\end{tabular} \\
		counterfactual network (C) & withtout Chugoku Expressway \\
		\hline
	\end{tabular}
\end{table}

To obtain inter-regional travel time data for the four states, we used the National Integrated Transportation Analysis System (NITAS) provided by the Ministry of Land, Infrastructure, Transport and Tourism (MLIT).
NITAS is a system that searches for routes between any domestic locations and calculates travel times and transportation costs for intermodal transportation networks that integrate road, rail, air, and sea transportation.

The study area encompasses the entire country of Japan and is classified into 207 zones, as defined in the National Trunk Passenger Flow Survey.
The origin and destination nodes of transportation are set at the location of the local government office of the city, where the population is the largest within each zone.

This study applies the SCGE model of \cite{ishikura2022} to estimate the general equilibrium state of the multi-regional economy corresponding to each transportation condition.
The model is based on the iceberg transportation cost concept, and inter-regional transportation travel times can be directly imported as input variables. 
The SCGE model explicitly treats a multi-industry economic system and is able to account for differences in trade barriers by sector.
We classify the industrial sectors into three categories (primary, manufacturing, and service) as in \cite{ishikura2022}.

\subsection{Effects in terms of transportation}
\label{transportation effects}
\begin{figure}
	\centering
	\includegraphics[width=0.75\linewidth]{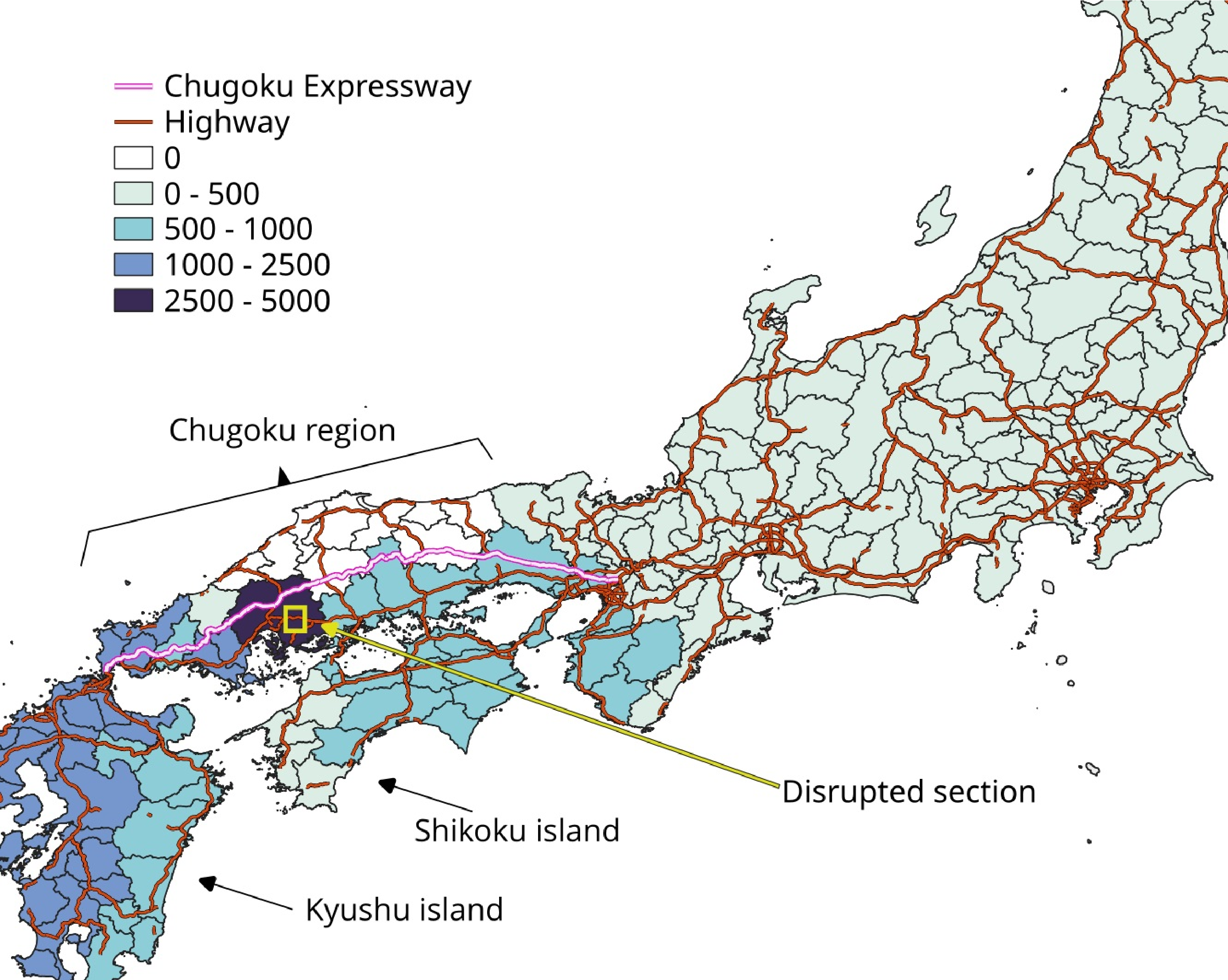}
	\label{fig:base time change}
	\\ (A-D) - (A-N) actual network

	\includegraphics[width=0.75\linewidth]{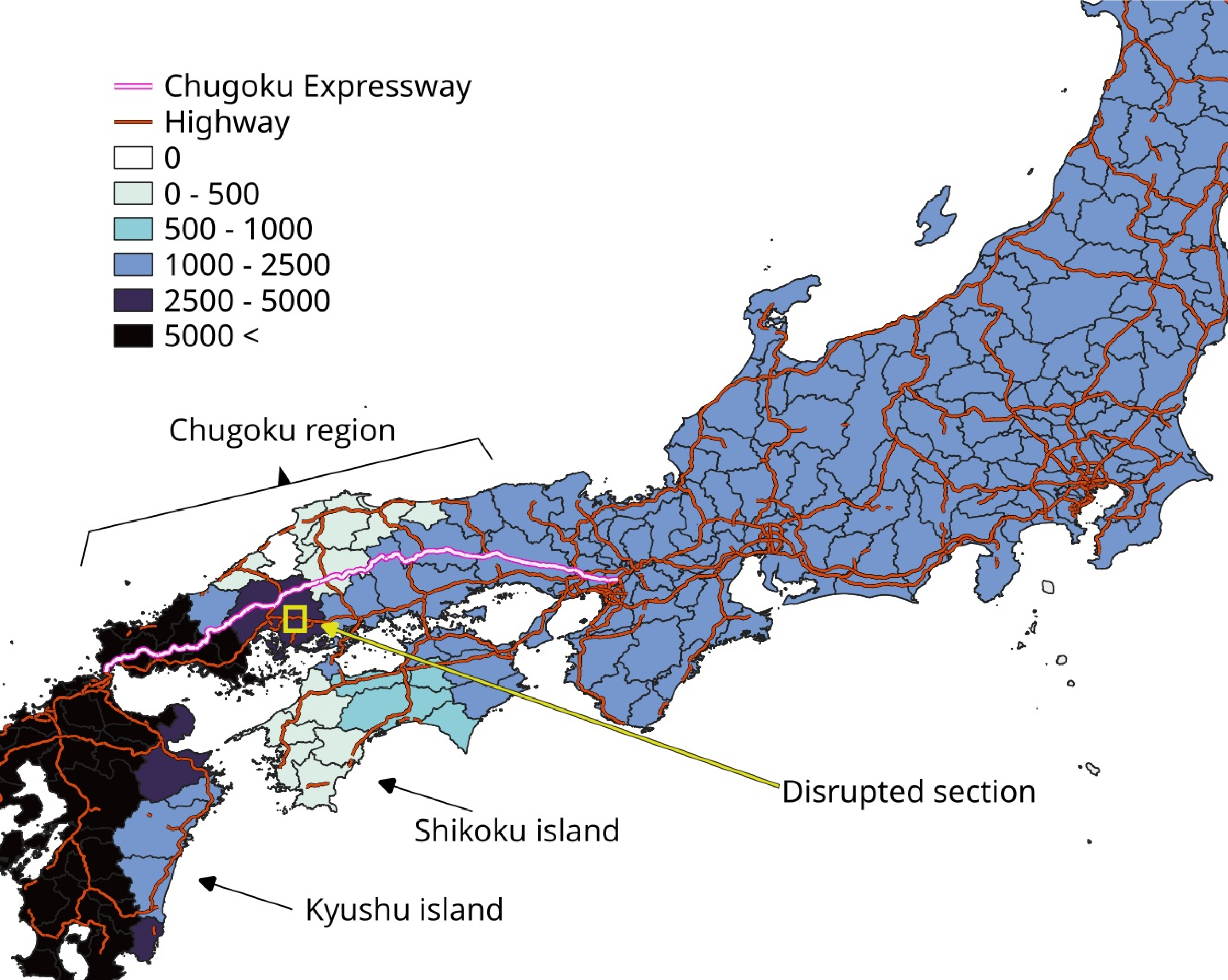}
	\label{fig:NoChugoku time change}
	\\ (C-D) - (C-N) counterfactual network
	\caption{Changes in shortest vehicle travel time aggregated on origin (unit: min)}
	\label{fig:travel time change}
\end{figure}

Fig. \ref{fig:travel time change} shows the impact of the highway disruptions on the A and C networks, respectively.
Although the length of the road disruption link is very small relative to the size of the national territory, the impact extends far from the affected area.
This indicates that the disrupted highways are part of the critical trunk roads for inter-city transportation.

In the actual road network (A), the effect is large in the vicinity of Hiroshima, which includes the disrupted section, and in the Kyushu region, located on the western side of the network.
On the other hand, in the counterfactual road network (C), the damage is particularly large in the Kyushu region, and the east side of the affected area is noticeably more affected than in the case of (A).
In the (C) network, the impact is much worse on the west side of the disrupted section, suggesting that the Chugoku Expressway was important as an alternative route, especially on Kyushu Island and the western edge of Mainland Japan.

\begin{figure}
	\centering
	\includegraphics[width=0.75\linewidth]{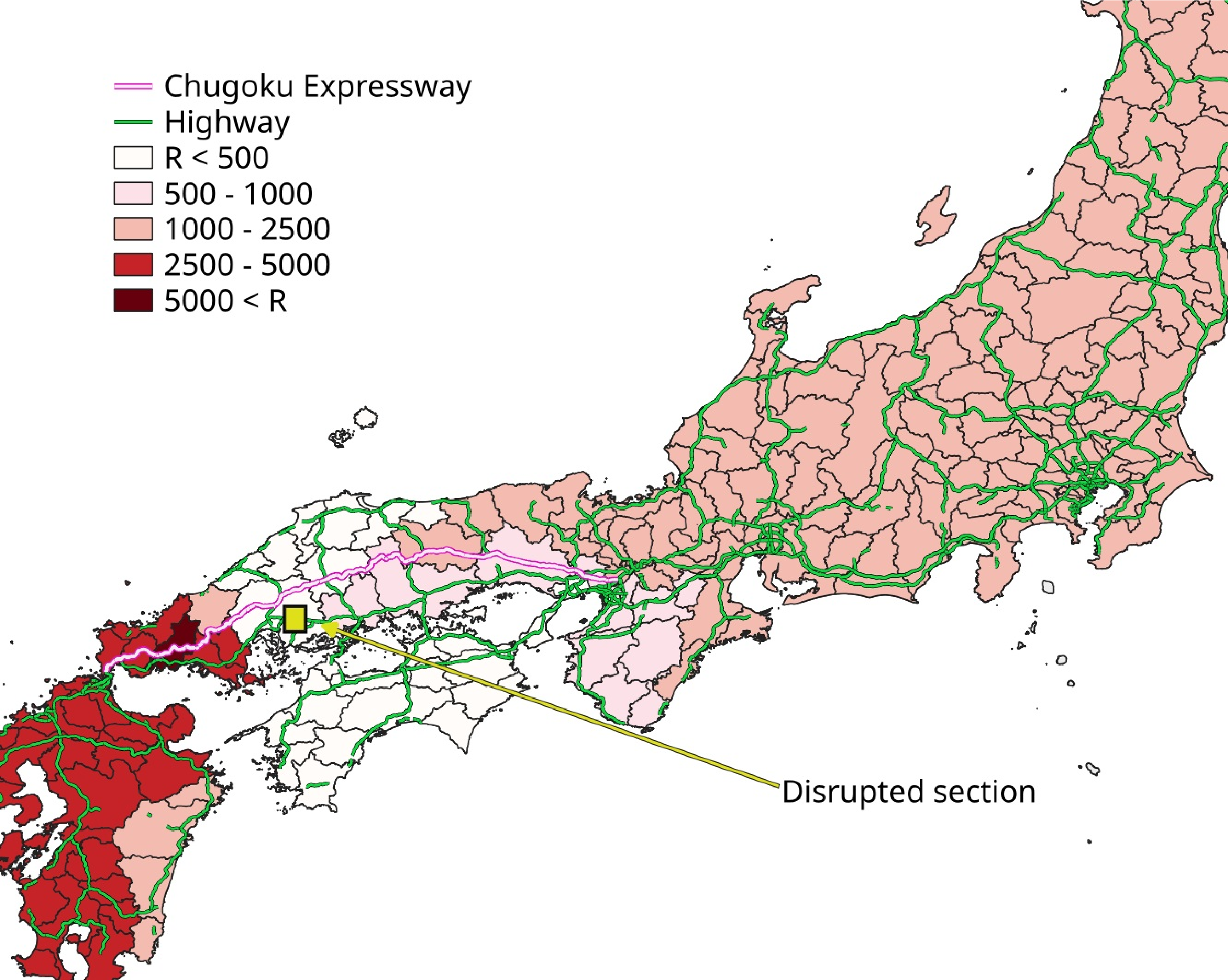}
	\caption{Effects of network redundancy in terms of transportation  (unit: min)}
	\label{fig:travel time reduncandy effect}
\end{figure}

\begin{figure}
	\centering
	\includegraphics[width=0.75\linewidth]{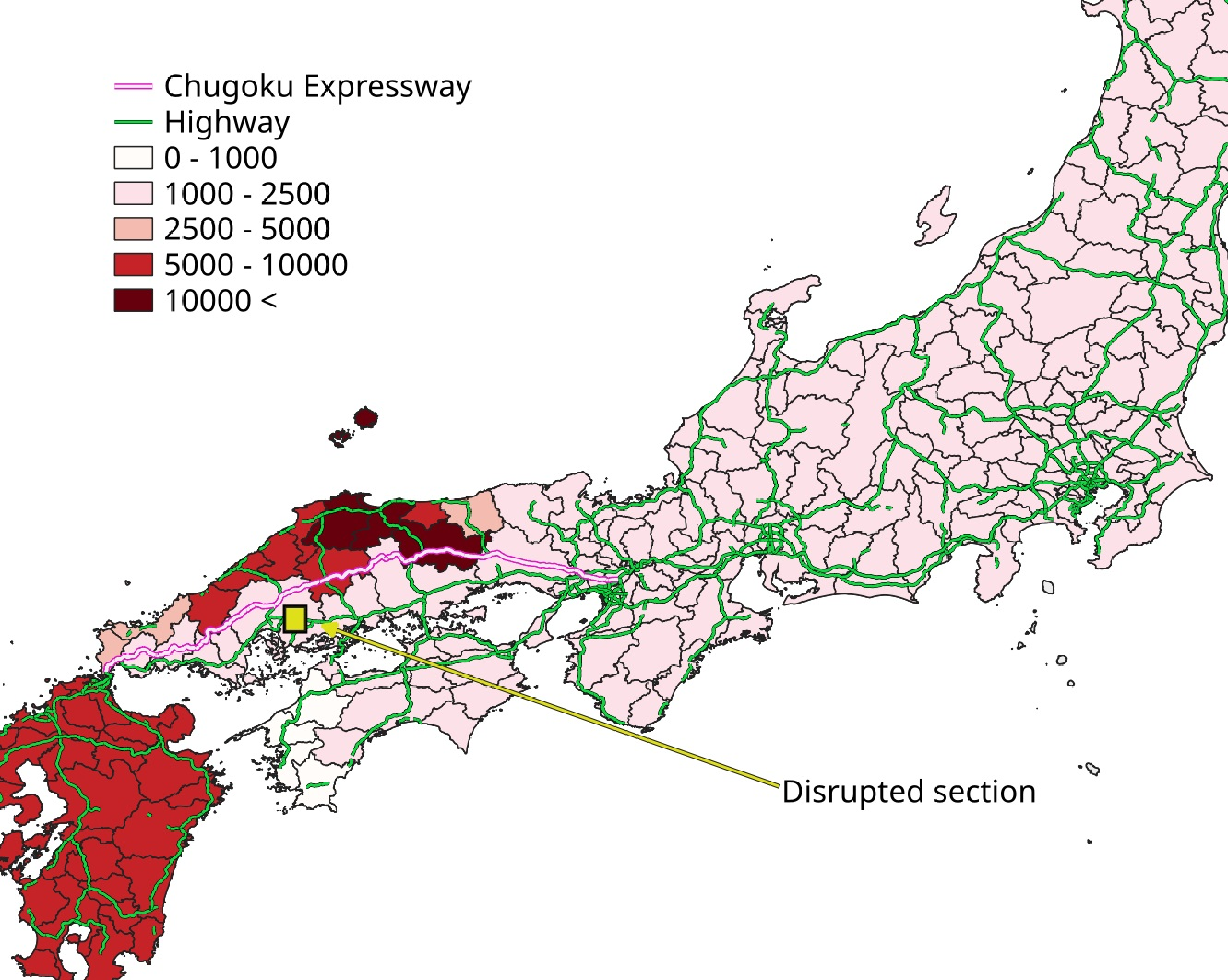}
	\caption{Travel time shortening (N, normal state, unit: min)}
	\label{fig:travel time direct effect}
\end{figure}

The difference between the two results, i.e., the network redundancy effect of the Chugoku Expressway, is shown in Fig. \ref{fig:travel time reduncandy effect}, which clearly shows the characteristics of the above context.
First, the effect of the redundancy network is large in the Kyushu Island region, especially in its western part.
On the other hand, the effect is smaller in Shikoku and San-in (north of the Chugoku region) than in the eastern half of Japan, despite their proximity to the directly affected areas.
This interesting feature requires additional discussion.
As the lower panel of Fig. \ref{fig:travel time change} shows, the northern part of the Chugoku region has a smaller impact on accessibility by the disruption of the affected section.
Thus, the disrupted sections (specifically, Hiroshima-Higashi - Shiwa) do not contribute much to the accessibility of transportation in these areas.
Instead, in the northern part of the Chugoku region, the accessibility improvement effect of the installation of the Chugoku Expressway is large (Fig. \ref{fig:travel time direct effect}).

For Shikoku Island, the accessibility effect of the Chugoku Expressway is small, as well as the network redundancy effect of the Chugoku Expressway.
In particular, when Hiroshima is hit by the disaster, the expressways in Shikoku Island can function as alternative routes, and the impact of the disaster on wide-area traffic is likely to be small.

\clearpage

\subsection{Effects in terms of welfare}
\label{economic effects}
The SCGE model depicts general equilibrium conditions in a multi-regional, multi-industry economic system.
The model outputs the price system, demand for goods (production), and income as the market equilibrium state, and the utility level is calculated from these values.
Here, the expenditure function in equilibrium, i.e., the utility level and the price system, are used to evaluate the welfare index EV described in the previous section.

\begin{figure}
	\centering
	\includegraphics[width=0.75\linewidth]{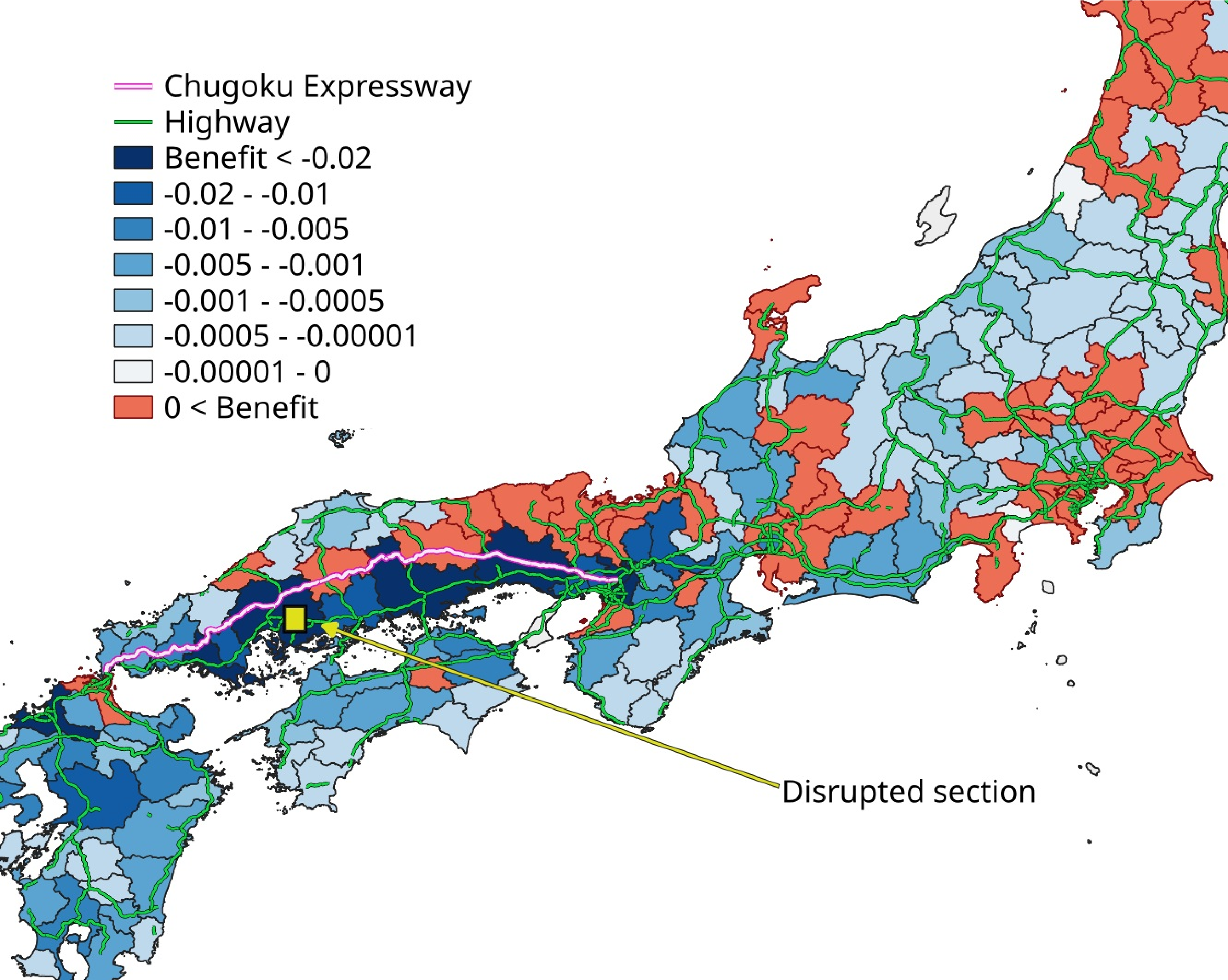}
	\label{fig:base EV}
	\\ (A-D) - (A-N) actural network

	\includegraphics[width=0.75\linewidth]{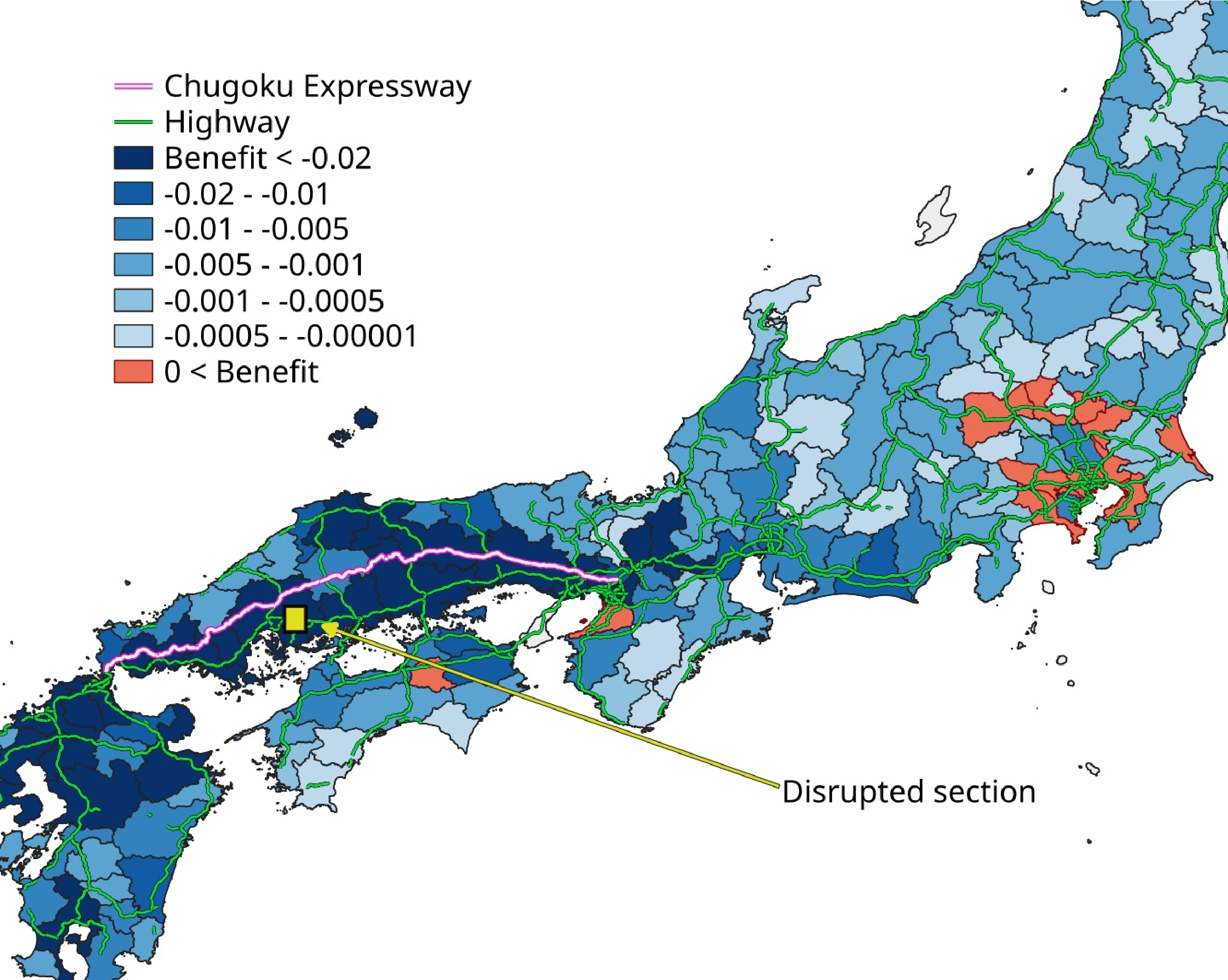}
	\label{fig:NoChugoku EV}
	\\ (C-D) - (C-N) counterfactual network

	\caption{Economic loss: measured by EV index (unit: trillion JPY)}
	\label{fig:compare EV}
\end{figure}

Fig. \ref{fig:compare EV} shows the regional welfare losses due to road disruptions in (A) the real network and (C) the counterfactual network, respectively.
The unit of EV is billion yen.
Unlike the impact measured by the travel time, there are some regions where the road disruption caused by a disaster has a positive benefit.
This characteristic is caused by the competitive relationship in the production of goods between regions where transportation accessibility is worsened by road disruption and regions where the impact of the disruption is small.
In other words, the increase in production costs associated with higher transportation costs raises the price of goods produced in the region and shifts demand to goods produced in other regions.
As a result, production in the region unaffected by the disaster increases, and factor incomes in that region also increase.

Geographically, in both cases (A) and (C), the welfare losses are larger in the southern part of the Chugoku region.
In case (C), where the Chugoku Expressway is excluded, negative benefits are observed in a larger number of areas, indicating a wider range of economic damage.

\begin{figure}
	\centering
	\includegraphics[width=0.75\linewidth]{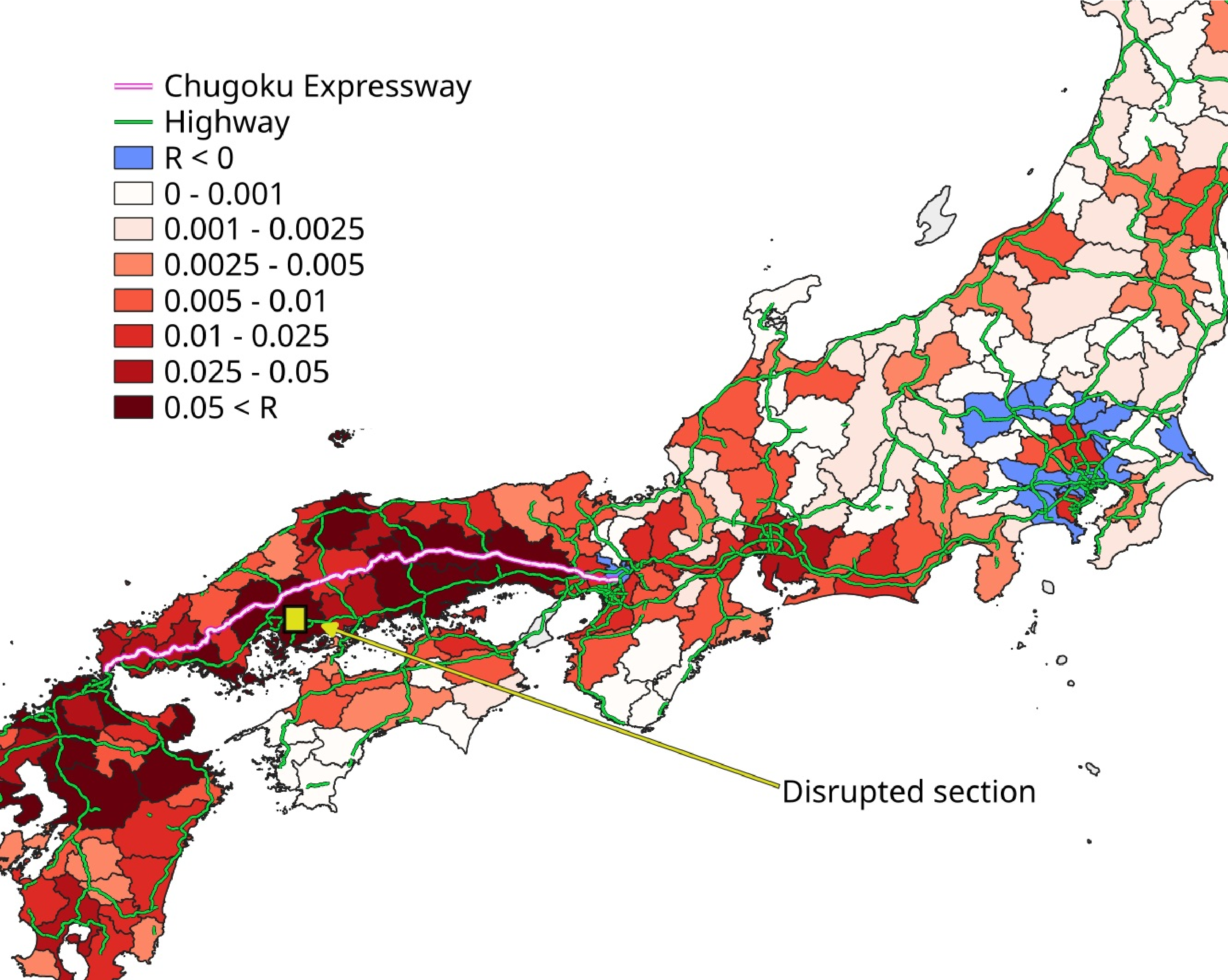}
	\caption{Effects of netrowk redundancy in terms of welfare (unit: trillion JPY)}
	\label{fig:EV redundandy effect}
\end{figure}

The regional distribution of these differences, the redundancy effect, is shown in Fig. \ref{fig:EV redundandy effect}.
The magnitude of the effect means how much economic vulnerability is mitigated by the Chugoku Expressway.
Compared to Fig. \ref{fig:travel time reduncandy effect}, which shows the effect in terms of transportation, there are many differences as well as similarities, such as the large effect in the Kyushu region.
The northern part of the Chugoku region and the areas east of the road disruption exhibit large effects.
The effect is also large in eastern Japan and along the Pacific coast of the central part of the Japanese Mainland.
The zones with large effects in the regions far to the east coincide with the core regions with large population size and gross regional product size in each area.
One of the main reasons why the regional characteristics of the network redundancy effect differ between the transportation and economic perspectives is the complexity of economic interdependence.

For example, consider the case where tradable goods from outside the region of production are purchased as intermediate inputs for production.
Even if no disaster-induced accessibility change occurs in the production region itself, if goods from the region with accessibility worsened are input, the cost of production will rise through an elevated price of goods produced in that region.
It is known as the forward linkage effect in the supply chain.
In addition, households' consumption expenditure also declines in regions where factor income declines due to a shrinkage in goods production.
The decrease in regional consumption expenditure also reduces the final demand for goods produced in other regions, which causes a decrease in the supply of those goods.
This is an interaction mechanism known as the backward linkage effect.
Therefore, depending on the input-output structure among regions and industries, the economic impact will spill over in a very complex manner.
Consequently, the regional distribution of economic impacts is more complex than that of transportation impacts.

\begin{figure}
	\centering
	\includegraphics[width=0.75\linewidth]{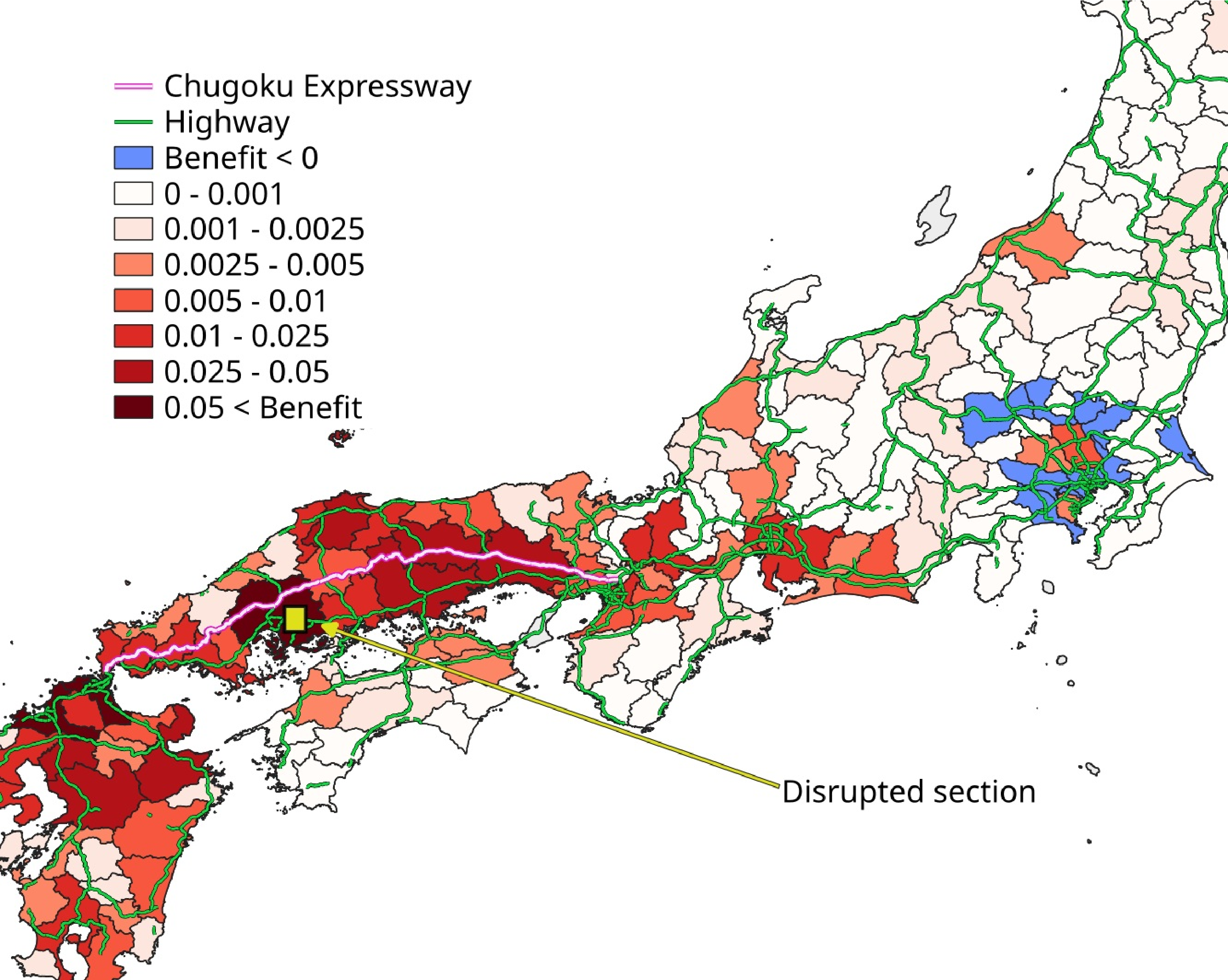}
	\caption{Distribution of benefit of Chugoku Expressway (unit: trillion JPY)}
	\label{fig:EV direct effect}
\end{figure}

Fig. \ref{fig:EV direct effect} shows the difference in welfare levels between the (C) and (A) networks at (N) with no disconnection $EV^{A-C,N}$, i.e., the distribution of the benefits of investment in the Chugoku Expressway.
\begin{equation}
	\label{investment EV}
	EV^{A-C,N} = e\left(U_{A,N},p_1^{A,N},p_2^{A,N}\right) - e\left(U_{C,N},p_1^{A,N},p_2^{A,N}\right)
\end{equation}
The project evaluation of highway investment generally considers only such benefits under normal conditions.
Fig. \ref{fig:EV direct effect} suggests that large benefits are attributed to the regions along and near both ends of the Chugoku Expressway.
However, the network redundancy effect (Fig. \ref{fig:EV redundandy effect}) is distributed over a wider area.
This implies that even regions where the direct economic benefits of travel time reduction are not large are benefiting from the reduction of economic damage in the event of a disaster.
These results quantitatively demonstrate the role of the Chugoku Expressway as an alternative route in the event of a disaster in Hiroshima.

Aggregated summary of welfare impacts is shown in Table \ref{tab: welfare}.

\clearpage

\begin{table}[t]
	\centering
	\caption{Aggregated summary of welfare impacts}
	\label{tab: welfare}
	\vspace{5mm}
	\begin{tabular}{cccccc}
		\hline
		zone name	&	zone ID	&	$EV^{A,N-D}$	&	$qEV^{C,N-D}$	&	$R$	&	$EV^{A-C,N}$	\\ \hline
		Tottori Prefecture	&	136-138	&	-0.17	&	-38.37	&	38.20	&	36.54	\\
		Shimane Prefecture	&	139-142	&	-1.23	&	-51.28	&	50.05	&	47.34	\\
		Okayama Prefecture	&	144,145	&	-39.40	&	-120.31	&	80.91	&	75.04	\\
		HIroshima-HIroshima	&	146	&	-125.73	&	-192.97	&	67.23	&	59.32	\\
		Hiroshima-Bingo	&	147	&	-17.77	&	-33.73	&	15.96	&	13.58	\\
		Hiroshima-Bihoku	&	148	&	0.17	&	-5.03	&	5.21	&	5.13	\\
		Yamaguchi Prefecture	&	149-154	&	-45.32	&	-174.18	&	128.85	&	74.67	\\
		Eastern	&	1-135	&	-97.52	&	-298.32	&	200.80	&	138.10	\\
		Shikoku-island	&	155-168	&	-32.96	&	-61.28	&	28.33	&	22.07	\\
		Kyushu-island	&	169-198	&	-88.23	&	-650.56	&	562.33	&	461.41	\\ \hline
		\multicolumn{6}{l}{unit: billion JPY}\\
		\multicolumn{6}{l}{See the appendix A. for the geographical location of zone ID.}
	\end{tabular}
\end{table}

\section{Concluding remarks}
\label{section: conclusion}
Road disruptions due to disasters can affect transportation accessibility in various areas, resulting in widespread economic inconvenience.
If effective alternative routes are available, the impact of road disruptions can be mitigated.
This reduction can be interpreted as an effect of the redundancy of the road network.
This study proposes a concept to quantitatively evaluate the contribution to network redundancy of targeted road segments.
A lot of existing studies have conducted similar analyses focusing on the transportation function of the road network.
In addition to this, we focused on the impact on economic conditions, which is a consequence of the road network condition.
We developed a method to measure the monetary value of the redundancy effect by comparing the equilibrium state of the economic system.

We further assessed the impact on regional welfare in Japan using the SCGE model, taking the actual road disruption case of the July 2018 Hiroshima torrential rain event.
The contribution of the Chugoku Expressway to network redundancy in terms of transportation was significant in the Chugoku and Kyushu regions, which are located close to the west side of the disrupted sections.
On the other hand, the network redundancy effect on the economic impact spread extensively to eastern Japan and these areas.

In the road investment project evaluation in Japan, only three types of benefits are measured for each invested road section: travel time reduction, travel cost reduction, and traffic accident reduction.
The benefits are obviously likely to be large for investment projects on roads with high traffic demand.
The Sanyo Expressway, which was the subject of the disruption scenario in our analysis, lies along regions of high economic activity and has high traffic demand.
Meanwhile, the Chugoku Expressway runs parallel to the Sanyo Expressway through mountainous areas.
Thus, the Chugoku Expressway, which has relatively small traffic demand, is less efficient than the Sanyo Expressway in terms of investment efficiency.
However, the Chugoku Expressway has great significance when the Sanyo Expressway is disrupted because it is a major alternative path to the Sanyo Expressway, which is the principal route between Kyushu Island and the center of Japan.
This study quantitatively demonstrates this corollary.
Our results shed light on the importance of network redundancy in areas exposed to disaster risk.

\section*{Acknowledgements}

This work was supported by the Kajima Foundation's Support Program for International Joint Research Activities, ``Innovation inspired by Nature'' Research Support Program, SEKISUI CHEMICAL CO., LTD. and JSPS KAKENHI Grann Number JP26K01067.

\clearpage

\appendix

\section{Regional classification}
\label{A_reigion classification}
\begin{figure}[h]
	\centering
	\includegraphics[width=0.9\linewidth]{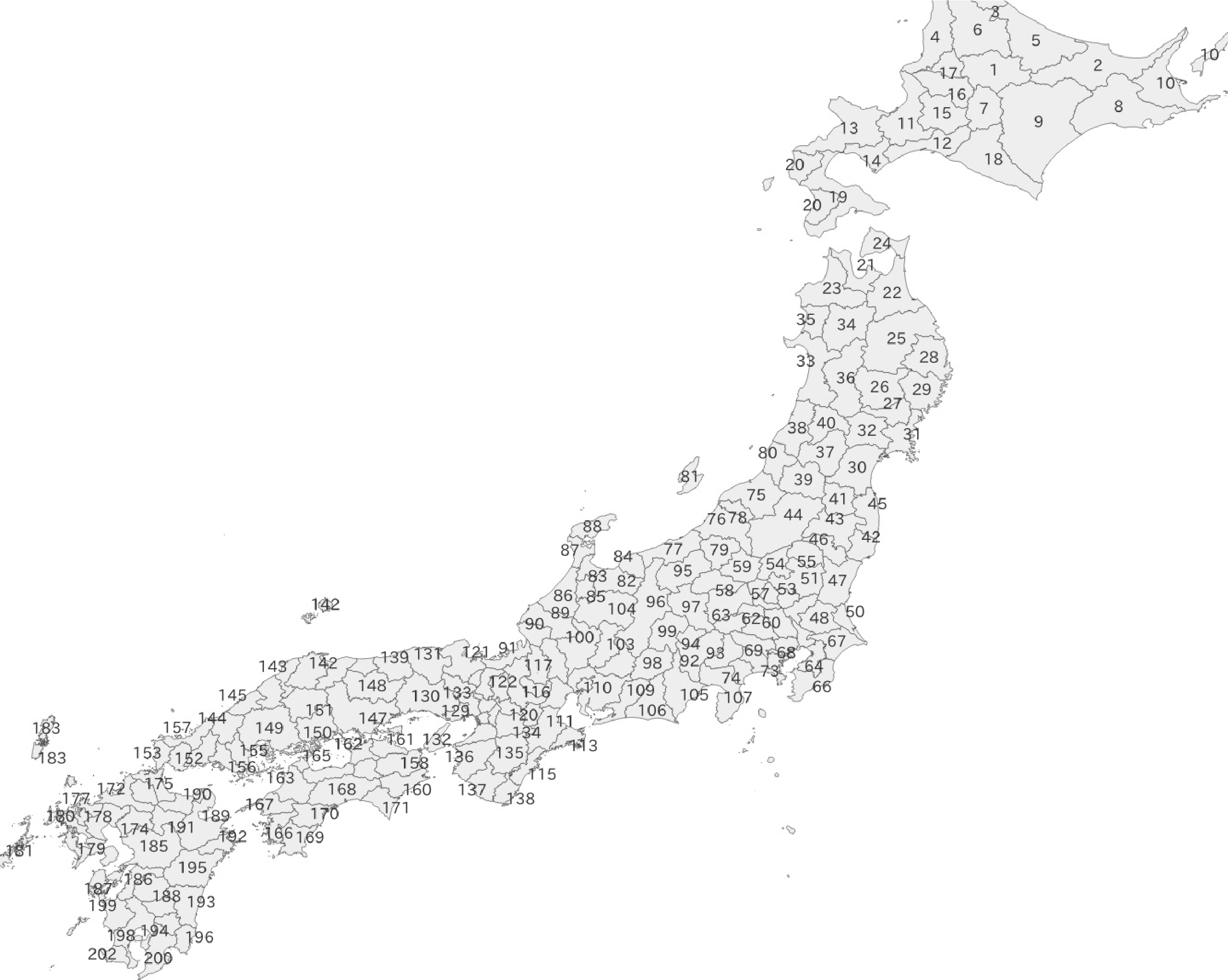}
	\caption{Definition of regional classification}
	\label{fig:region classification}
\end{figure}

The locations and ID of each zone correspond as shown in Fig \ref{fig:region classification}.

\bibliographystyle{plainnat}

\bibliography{Ishikura2025Aug}

\end{document}